\documentstyle[12pt]{article}
\author{Yonathan Shapir$^{1,2}$ and Serge Galam$^2$,\\ 
$^1$Department of Physics and Astronomy,\\
University of Rochester, Rochester, New York 14627.\\
$^2$Groupe de Physique des Solides\footnotemark[1],\\
Tour 23, 2 place Jussieu, 75251 Paris Cedex 05, France.\\[1ex]}
\title{Continuous Versus First Order Transitions \\in \\Compressible Diluted 
Magnets}
\date{ Physica A, \underline{224}, 669, 1996\\
\begin{center}
{\em PA Classification Numbers:\/} 05.50, 64.70, 75.10H \\[6ex]
\end{center}
}
\addtocounter{footnote}{0}
\footnotetext{Laboratoire associ\'{e} au C.N.R.S. (U.A. n$^{\circ}$
 17) et
aux Universit\'{e}
Paris VII et Paris VI.}
\addtocounter{footnote}{1}

\begin{document}
\maketitle
\baselineskip 3.3ex
\footskip 5ex
\parindent 2.5em
\abovedisplayskip 5ex
\belowdisplayskip 5ex
\abovedisplayshortskip 3ex
\belowdisplayshortskip 5ex
\textfloatsep 7ex
\intextsep 7ex

\begin{abstract}
The interplay between disorder and compressibility in Ising magnets
 is studied.
Contrary to pure systems in which a weak compressibility drives the
transition first order,
we find from a renormalization group  analysis that it has no effect
 on
disordered systems
which keep undergoing  
continuous transition with rigid random-bond Ising model critical
exponents. The mean field
calculation exhibits a dilution-dependent tricritical point beyond
 which,
at stronger compressibility
the transition is first order. The different behavior of XY and Heisenberg
magnets is 
discussed.

\end{abstract}
\baselineskip 24pt
\newpage
\section{Introduction}
The effects of the magneto-elastic coupling (MEC), between the spin
 and the
lattice vibrational
degrees of freedom, have been investigated extensively [1-10]. Basic
thermodynamics implies that for magnetic systems with a diverging
 specific
heat the
presence of the MEC will cause an instability. Therefore the divergence
 is
averted
by a preemptive first-order transition. Mean-field theory (MFT) and
solutions of simplified models
[3-6] have vindicated this conclusion. In particular
MFT predicts the existence of a tricritical point separating second-
 and first-
order transitions, as the intensity of the MEC is increased [5].
 Other types of behavior were predicted upon changing
boundary and other conditions. However, they will not be discussed
 here [11].
An even better understanding was achieved with the application of
 the
renormalization
group (RG) approach to these systems [7-9]. Calculations in d=4-$\epsilon$
dimensions,
have shown the behavior to depend crucially on the number of components
 m of the
order-parameter. For example the ferromagnetic continuous transition of
rigid Ising-like
systems (m=1) is fluctuation-driven to a first-order transition
once the MEC is accounted for. 
On the other hand XY (m=2) and Heisenberg (m=3) systems are unaffected
by the compressibility and the MEC is irrelevant [7-9]. This particular
aspect is in
accordance with the early results since the crossover exponent for the MEC
is $\alpha$,
the rigid system specific heat exponent and the interaction is relevant for
rigid systems with a divergent specific heat, i.e. $\alpha>0$.

Experimentally, the coupling is exhibited by the sensitivity of the
 critical
temperature to an applied external pressure. First-order and continuous
transitions, in some cases separated by a tricritical point, were
 reported
in different
systems [12-16].

In the present article we shed a new light on the behavior of compressible
magnets.
We emphasize the drastic role played by quenched bond disorder, such
 as dilution
(obtained by replacing some of the ions by non-magnetic ones). We
 find that for
Ising-like systems the quenched disorder has a crucial effect in
 preventing the
runaway to the first-order transition caused by the MEC in pure systems.
This drastic effect of the disorder originates from the so-called
Harris criterion: Weak disorder is relevant for rigid systems with
positive $\alpha$ (namely precisely the same systems that are affected
by the MEC). These disordered rigid systems exhibit a new critical behavior
associated with a random fixed-point for which $\alpha<0$. Hence
one possibility is that a weak MEC will not affect their behavior
(another possibility being that the MEC has the dominating effect: 
It drives the transition 
discontinuous which is then somewhat weakened by the disorder).
In the next Section (2) the former heuristic
conclusion is confirmed by a detailed RG analysis.
 In Sec. 3 the MFT is applied 
and its results are compared with those 
 obtained by the same method for pure compressible magnets. 
In the last Section (4) the important conclusions are summarized and
discussed within the context of other bond-disorder effects 
in low-dimensional magnetic models.  

\section{Renormalization group analysis}
The momentum-shell RG analysis begins from a continuous-spin Landau-
Ginzburg-Wilson Hamiltonian. With $\vec{\phi}(\vec{x})$ representing
 the
m-component
spin field and $\vec{u}(\vec{x})$ the d-dimensional local displacement
 the
Hamiltonian
is [4,7]:
\begin{eqnarray}
       -\beta H =
                    &   & \int
d^dx\{\frac{1}{2}r(\vec{x}){(\vec{\phi}(\vec{x})})^2+
\frac{1}{2}\sum_{a=1}^{m}{(\vec{\nabla }\phi
_a(\vec{x}))}^2+u_0{(\vec{\phi}(\vec{x}))}^4+
 \\
                    &   &
(\frac{1}{2}K-\frac{1}{d}\mu){(\vec{\nabla}\cdot\vec{u}(\vec{x}))}^2+
 \mu \sum_{\nu=1}^d{(\vec{\nabla}u_{\nu}(\vec{x}))}^2+ 
g({\vec{\nabla }\cdot\vec{u}){(\vec{\phi}(\vec{x}))}^2}\}\,. \nonumber
\end{eqnarray}
   The first three terms represent the rigid random-bond Ising model.
 The
dilution is accounted for in $r(\vec{x})=r+$$\delta$$r(\vec{x})$,
 where r
is the average
of the local random ``temperature" $r(\vec{x})$. The local deviations
 obey
$$
<\delta r(\vec{x}) \delta r(\vec{y})>=<\delta r^2>\delta (\vec{x}-\vec{y})
\,,
$$
(higher moments are irrelevant and so are the omitted disorder-dependent
terms in the elastic part of the Hamiltonian). 
K and $\mu$ are the bulk and shear moduli and g
is the MEC. To average over the disorder we use the ``n goes to zero"
replica trick.
After replicating and averaging the local displacements may be integrated
 out
in $<Z^n>$. The outcome is an effective replicated Hamiltonian depending
only the
n m-component vector $\vec{\phi}^\alpha$, $\alpha$=1,2,....,n:
 \begin{eqnarray}
-\beta\overline {H_{rep}}=
                             &   & \int d^dx\{ \sum_{\alpha }
(\frac{1}{2}r(\vec{\phi }^{\alpha })^2
+\frac{1}{2}\sum_{a=1}^{m}(\vec{\nabla }\phi _a ^{\alpha })^2+u(\vec{\phi
^{\alpha }})^4)\} \\
                             &   & +\sum_{\alpha } \frac{v}{V}(\int
 d^dx
(\vec{\phi ^{\alpha }})^2)^2
-\triangle \int d^dx (\sum_{\alpha }(\vec{\phi ^{\alpha }})^2)^2
 \,, \nonumber
\end{eqnarray}
where:
$$
u=u_0-\frac{g^2}
{ 4 \{ \frac{1}{2} K+ ( \frac{d-1}{d}) \mu \} }\,, \nonumber
$$
$$
v=\frac{1}{4}g^2- \frac{1}{ \{ \frac{1}{2} K+(\frac{d-1}{d}) \mu
 \}
}+\frac{2}{K}\,,
$$
and
$$
\Delta =8<\delta  r^2>\,.
$$
The initial (and physically acceptable) values obey $u>0$, $\Delta>0$,
 $v<0$.
V is the volume of the system.

The Hamiltonian has three quartic couplings. The flow within the
 three
dimensional
subspace of parameters which contains u, $\Delta$, and $v$ determines
 the
fixed-point
structure of the system. Integrating out small scales fluctuations,
 we
obtain the
recursion relations to order one loop:
\begin{equation}
\frac{\partial u}{\partial l}=\epsilon u-\frac{m+8}{6}u^2+2\Delta
 u\,,
\end{equation}
\begin{equation}
\frac{\partial \Delta}{\partial l}=\epsilon \Delta+\frac{mn+8}{6}\Delta
^2-\frac{m+2}{3}\Delta u\,,
\end{equation}
\begin{equation}
\frac{\partial v}{\partial l}=\epsilon
v-\frac{m}{2}v^2-\frac{m+2}{3}uv+\frac{n+2}{3}\Delta v\,,
\end{equation}

One observation is called for at once: the coupling $v$ cannot feed
 into
the renormalization
of the two others because of its special non-local form. Consequently
  we
separate the search
for the fixed points (f.p.) into two steps: We look first for the
 f.p. in
the u-$\Delta$
directions. Secondly, with these f.p. values for u and $\Delta$
inserted into Eq. (5) we look for its fixed-points. The first step
 is
analogous to
the analysis done in the study of rigid random-bond models. The results
 are also
m-dependent [17]. For the Ising model the pure f.p. is unstable towards
 the
inclusion of the
disorder coupling $\Delta$, while it is stable for systems with $m>1$
[17,18]. We therefore discuss
the results for these cases separately.

\subsection{ Ising model (m=1)}

   The stable f.p. in the u-$\Delta$ plane is the so-called Khmel'nitskii's
f.p. [18] at which
   both u and $\Delta$ are of order $\sqrt{\epsilon}$. We therefore
 have to
check the
   stability of this f.p. while switching on v. To that goal we insert
 the
values of the f.p. $u^*=\sqrt{\frac{6}{53}}\epsilon ^{\frac{1}{2}}$
 and
   $\Delta ^*=\frac{3}{4}u^*$
  to Eq. (5). These values were obtained from two-loop
   calculations [18,19].
For $v$ close to zero the term in $v^2$ is negligible and thus we
 have:
\begin{equation}
\frac{\partial v}{\partial l}=(\epsilon -u^*+\frac{2}{3}\Delta
^*)v=-2\sqrt{\frac{6}{53}}
\epsilon ^{\frac{1}{2}}v\,.
\end{equation}

   Terms of order $\epsilon$ were neglected compared with $\sqrt{\epsilon}$.
It should be noticed that the crossover exponent for $v$ is $\alpha$ of the
rigid random-bond system. This is contrary to the case in which a random-field
is applied to the random-bond system and for which the crossover exponent is
not $\gamma$ of the bond-disordered model [20]. Indeed, in the present case
the $v$ term is the square of the internal energy (per unit volume)
and hence has $\alpha$ as its scaling exponent. 
   Our finding $dv/dl<0$ implies that this fixed point is  the stable  one
and hence
   is describing the critical behavior of this system (at least within
 the
renormalized
   perturbation theory which starts from weak couplings). We thus
 conclude
that elastic random-bond Ising systems will exhibit, at their critical point, the critical
exponents of the
   rigid random-bond Ising system. If the MEC coupling $v$ is large and that
of the disorder $\Delta$ is small, the initial flow will be towards larger $v$.
The stability boundary may be reached before the flow reverses itself to smaller
values. In this case a first-order transition is still possible and the continuous
transition will be attained (through a tricritical point) at a finite strength of
the disorder.
 That is very different from the behavior
in absence
   of disorder for which it was found that the elasticity is relevant
 at
the order
   $\epsilon$ Wilson-Fisher f.p. (of the pure-rigid model [7,8])
 and drives
the transition
   first-order. The flow diagram is schematically depicted in Fig.
 1.

\subsection{Continuous spin models $(m>1)$: XY, Heisenberg, etc.}

   For these models with continuous symmetry of the order parameter
 the
stable f.p.
   in the u-$\Delta$ plane is that of the pure system. We therefore
 have to
insert
   the values at this f.p. $u^*=\frac{6\epsilon}{m+8}$  [17] into Eq. (5):

\begin{equation}
\frac{\partial v}{\partial l}=(\epsilon-\frac{6(m+2)}{3(m+8)}\epsilon)v
=(\frac{4-m}{m+8})\epsilon v\,,
\end{equation}
   Although $dv/dl$ is positive for $m<4$ and small $\epsilon$ at the $v=0$
 f.p., it is again given by $\alpha$ of the rigid (and pure for the
considered) systems. As discussed above this is a general relation 
which holds to all orders in $\epsilon$. It is known [17] that
$\alpha$ becomes negative for larger values of $\epsilon$ and is
definitively so at $\epsilon=1$ for all $m>1$.
Therefore $dv/dl$ is negative in 3d and
the same f.p. of
   the pure and rigid systems also describe their critical properties
 in
   presence of both disorder and compressibility (and their exponents
 remain
   unchanged).
 
\section{Mean Field Theory} 
To go beyond the renormalized perturbation expansion we need to have
 recourse
to mean-field theory (MFT), although its results are always to some
 extent
questionable (e.g. they are insensitive to the dimensionality of
 the
embedding space).
However, we may gain some insight by comparing
within the same approach the predictions for the critical behavior
 with and
without
dilution.

MFT is applied to the lattice model with $\vec{S}^2=1$ on every
 site. We choose
a somewhat simplified (Domb's) model which was shown [3, 5] to
capture the essential physics. In this model all nearest-neighbor
 exchange
interactions are strengthened (or weakened) by the same amount depending
 on
the average inter-site distance. The deviation in this distance is
 proportional
to the change in the volume fraction $w=V/N$ ($N$ is the
number of sites) from its average $w_0$.

The Hamiltonian contains two contributions: the elastic and the magnetic.
 The
elastic part is:
 \begin{equation}
H_{el}=\frac{N\phi}{2}(w-w_0)^2\,,
\end{equation}
where $\phi$ is proportional to the elastic constant. The magnetic
part is [3]:
\begin{equation}
H_{mag}=\sum_{i,j}J_{i,j}(w)S_iS_j\,,
\end{equation}
where we have chosen $S_i$ to be a scalar (Ising spin) keeping in
 mind that MFT
yields similar results for m-vector spins.
The nearest-neighbor couplings are given by:
\begin{equation}
J_{i,j}(w)=\epsilon _{i,j} \{J_0+J_1(w-w_0)\}\,,
\end{equation}
$\epsilon _{i,j}$ are independent random variables with the distribution
$p(\epsilon )=x\delta (\epsilon )+(1-x)\delta (\epsilon -1)$, where
 $x$ is
the concentration
of non-diluted
bonds (similar results will be obtained for either site or bond dilution).
$J_1$ is the coefficient of the part  which depends on the inter-site
 distance
(only the dominant, linear, dependence has been kept).

It is straightforward to integrate out the volume fluctuations. The
remaining effective Hamiltonian depends only on the spin degrees
 of freedom:
\begin{equation}
-\beta H_{eff}=\beta J_0\sum_{i,j}\epsilon _{i,j}S_iS_j+
\frac{\beta J_1^2}{2N\phi}{(\sum_{i,j}\epsilon _{i,j}S_iS_j)}^2\,.
\end{equation}
Introducing mean-fields, both terms are reduced to single-spins ones.
Expanding the free energy and
performing the average over the disorder yields the following effective
mean-field
Hamiltonian:
\begin{equation}
-\beta \overline { H_{MF}}=\beta J_0xcm\sum_{i}S_i-\frac{N}{2}\beta
 J_0 xcm^2+
\beta \frac{J_1^2}{4\phi}x^2c^2m^3\sum_{i}S_i-\frac{N}{8\phi}\beta
 J_1^2
x^2c^2m^4\,,
\end{equation}
where c is the coordination number of the lattice and here $m$ is
 the
magnetization per site.
>From this Hamiltonian all thermodynamic properties can be calculated.
 In
particular for the averaged free-energy we obtain:
\begin{eqnarray}
-\beta \overline {f_{MF} } & = & \lim_{N\rightarrow \infty } \overline
{=\frac{1}{N} lnZ }  \\
                           & = &
-\frac{\beta J_0 xcm^2}{2}-\frac{\beta J_1^2 x^2c^2m^4}
{8\phi}+ln\,cosh\{{\beta J_0xcm+\frac{\beta {J_1x^2c}^2m^3}{4\phi}}\}
 \,.\nonumber
\end{eqnarray}
The critical temperature is given by that at which the $m^2$ coefficient
vanishes: $kT_c=xcJ_0$. The transition becomes discontinuous when
 the $m^4$
coefficient becomes negative. Therefore the tricritical point is
 at
$kT_c=xcJ_0$ and $J_1^2=\frac{2J_0^2 \phi}{3kT_c}$.
We see first that within MFT the critical temperature depends only
 on
the dilution with the same dependence as that of the rigid system.
 The
value of $J_1/J_0$ beyond which the transition turns first-order
 does
depend on the dilution: The more diluted is the system the larger
 this
parameter must be in order to change the order of the transition.
 Hence
the regime of second-order is enlarged at the expense of the first-order
one the more diluted the system is. The phase diagram is drawn in
 Fig. 2.
This trend is consistent with the RG results for
the Ising systems in which a first-order transition was turned second
order by the dilution.

\section{Conclusions}
   To summarize, we have shown that disorder (e.g. dilution) 
has a crucial role in
compressible Ising magnets: While pure systems are driven first-order
by coupling spin fluctuations to the elastic degrees of freedom,
 none
of that happens in disordered systems. Their critical properties are
insensitive to the
compressibility and continue to exhibit a random-bond continuous
transition. This behavior is consistent with that of the specific-heat.
Indeed, the Harris's [21] and other [17,22] arguments imply that
 $\alpha$
of the pure system
is also the crossover exponent for the dilution, which is therefore
relevant for Ising systems (and irrelevant for $m>1$ vector-models).

Consequently, Chayes et al. [23] have shown that in any random system
 $\alpha$
is negative. Hence we reach the general conclusion that the compressibility
cannot be relevant at a second-order transition of a random system.
Our conclusions are also consistent with those of Refs. [24]. There it was shown
that bond-disorder suppresses (in $d=3$) any
tricritical point (it eliminates it in $d=2$ and therefore 2d
compressible magnets undergo always a continuous transition in the presence of 
disorder).

Although the crossover from one universality class to another is
slow ($\alpha$ being a small exponent), compressibility and bond-disorder
are present in many physical systems. 
 Therefore it will be very worthwhile 
to observe experimentally the effects of dilution 
in particular on the transition
in compressible magnets with uniaxial anisotropy.

\subsection*{Acknowledgments}
We are thankful to A. Aharony and D.J. Bergman for helpful comments.
One of us (Y.S.) is grateful for the hospitality of the Groupe de
Physique des Solides at Universit\'{e} Paris 7. He also acknowledges
 the
Donors of the Petroleum Research Fund administrated
by the ACS, for
 partial
support of this research. 


\newpage
\begin{center} {\bf REFERENCES}
\end{center}
$[1]$ O.K. Rice, J. Chem. Phys. {\bf 22}, 1535 (1954).\\
$[2]$ C. Domb, J. Chem. Phys. {\bf 25}, 783 (1956).\\
$[3]$ D.C. Mattis and T.D. Schultz, Phys. Rev. {\bf 129}, 175 (1963).\\
$[4]$ A.I. Larkin and S.A. Pikin, Zh. Eksp. Teor. Fiz. {\bf 56}, 1664 (1969) [Sov.

    Phys. JETP {\bf 29}, 891 (1969)].\\
$[5]$ S. Salinas, J. Phys. C7, C{\bf20} (1974).\\
$[6]$ V.B. Henriques and S. Salinas, J. Phys. C{\bf20}, 2415 (1987).\\
$[7]$ J. Sak, Phys. Rev. B{\bf10}, 3957 (1957); J. Bruno and J. Sak
 {\it
ibid} {\bf 22},
3302

 (1980).\\
$[8]$ F.J. Wegner, J. Phys. C{\bf 7}, 2109 (1974).\\     
$[9]$ D.J. Bergman and B.I. Halperin, Phys. Rev. B{\bf 13}, 2145
 (1976).\\
$[10]$ A.F.S. Moreira and W. Figueiredo, Phys. Rev. B{\bf 46}, 2891
 (1992).\\
$[11]$ See, e.g., Ref. [9] for a comprehensive overview.\\
$[12]$ C.W. Garland and R. Renard, J. Chem. Phys. {\bf 44}, 1130
 (1966); B.B.

 Weiner and C.W. Garland,
{\it ibid} {\bf 56}, 155 (1972); C.W. Garland and R.J.

 Pollina {\it ibid} {\bf 58}, 5502 (1973).\\
$[13]$ C.W. Garland and B.B. Weiner, Phys. Rev. B{\bf 3}, 1634 (1971).\\
$[14]$ C. P. Slichter {\it et al.}, Phys. Rev. B{\bf 4}, 907 (1971).\\
$[15]$ E. Erwin {\it et al.}, Physica B{\bf 161}, 260 (1989).\\
$[16]$ H. Yoshida {\it et al.}, J. Magn. Magn. Mater., {\bf 90 +
 91}, 177
(1990).\\
$[17]$ A. Aharony in {\it Phase Transitions and Critical Phenomena},
 Vol. 6,\\

 p. 358 (C. Domb and M.S. Green - Edit., Academic Press, N.Y. 1976).\\
$[18]$ D.E. Khmel'nitskii, Zh. Eksp. Teoret. Fiz. {\bf 68}, 1960
 (1975) [Sov.

 Phys. JETP {\bf 41}, 981 (1976)].\\
$[19]$ G. Grinstein and A. Luther, Phys. Rev. B{\bf 13}, 1329 (1976).\\
$[20]$ A. Aharony, Euro. Phys. Lett. {\bf 1}, 617 (1986).\\
$[21]$ A.B. Harris, J. Phys. C{\bf 7}, 1671 (1974).\\
$[22]$ Y. Shapir and A. Aharony, J. Phys. C{\bf 14}, L905 (1981).\\
$[23]$ J.T. Chayes {\it et al.}, Phys. Rev. Lett. {\bf 57}, 2999
 (1986).\\
$[24]$ M. Aizenmann and J. Wehr, Phys. Rev. Lett. {\bf 21}, 2503 (1989);

   K. Hui and A.N. Berker, {\it ibid}, 2507 (1989).
\newpage
\begin{center} {\bf FIGURES CAPTIONS} \\

\end{center} 

Fig. 1.   The three-dimensional flow diagram with the fixed points:
  \\
\hspace*{2.3cm}  (A) Pure-Rigid and (B) Disordered-Rigid.

Fig. 2.   Schematic mean-field phase diagram ($Y\equiv \frac{J_1^2}{\phi
}$).   
\end{document}